\documentclass[11pt]{article}
\usepackage{times}
\usepackage{fullpage}
\usepackage{amsmath}
\usepackage{amssymb}
\usepackage{amsfonts}
\usepackage[all,poly]{xy}

\newcounter{algoctr}

\newif\ifnotesw\noteswtrue
   {\ifnotesw\marginpar[\hfill\(\top\)]{\(\top\)}\fi}%
   {\ifnotesw\marginpar[\hfill\(\bot\)]{\(\bot\)}\fi}

\newcommand{\mnote}[1]%
    {\ifnotesw\marginpar%
        [{\scriptsize\begin{minipage}[t]{\marginparwidth}
        \raggedleft#1%
                        \end{minipage}}]%
        {\scriptsize\begin{minipage}[t]{\marginparwidth}
        \raggedright#1%
                        \end{minipage}}%
    \fi}

\newcommand{\ignore}[1]{}

\newsavebox{\given}
\savebox{\given}[1em]{\rule[-1.5ex]{.2mm}{4ex}}

\newcommand{\bnum}{\begin{equation}}
\newcommand{\enum}{\end{equation}}

\newtheorem{theorem}{Theorem}

\newtheorem{claim}[theorem]{Claim}
\newtheorem{conjecture}{Conjecture}

\newtheorem{definition}{Definition}

\newcommand{\blackslug}{\rule{7pt}{7pt}}

\newcommand{\real}{\ifmmode {\rm R} \else ${\rm R}$ \fi}
\newcommand{\nat}{\ifmmode {\rm N} \else ${\rm N}$  \fi}
\newcommand{\tot}{\ifmmode {\cal T} \else ${\cal T}$ \fi}
\newcommand{\sigstar}{\ifmmode \Sigma^{\ast} \else $\Sigma^{\ast}$ \fi}

\newcommand{\inn}{\ifmmode \in \else $\in$ \fi}
\renewcommand{\phi}{\ifmmode \varphi \else $\varphi$ \fi}
\renewcommand{\le}{\ifmmode \leq \else $\leq$ \fi}
\renewcommand{\ge}{\ifmmode \geq \else $\geq$ \fi}
\renewcommand{\ne}{\ifmmode \neq \else $\neq$ \fi}
\newcommand{\lt}{\ifmmode < \else $<$ \fi}
\newcommand{\gt}{\ifmmode > \else $>$ \fi}
\newcommand{\eq}{\ifmmode = \else $=$ \fi}
\newcommand{\half}{\ifmmode \frac{1}{2} \else $\frac{1}{2}$ \fi}
\newcommand{\oneovern}{\ifmmode \frac{1}{n} \else $\frac{1}{n}$ \fi}

\newcommand{\ra}{\ifmmode \rightarrow \else $\rightarrow$ \fi}

\renewcommand{\notin}{\ifmmode \not\in \else $\not\in$ \fi}
\newlength{\thislabel}
\newcommand{\labsize}[1]{\settowidth{\thislabel}{#1}}

\newcommand{\lip}{\langle}
\newcommand{\rip}{\rangle}

\def\Int{\mathbb Z}
\def\Real{\mathbb R}

\def\dotdef{\stackrel{\cdot}{=}}

\newcommand{\ket}[1]{| #1 \rip}
\newcommand{\braket}[2]{\lip #1 | #2 \rip}
\newcommand{\tderiv}[1]{\frac{\mathsf{d}}{\mathsf{d}t} #1}

\title{{\sf Mixing in Continuous Quantum Walks on Graphs}}
\author{
{Amir Ahmadi}
\and
{Ryan Belk}
\and
{Christino Tamon}\footnote{Contact email: tino@clarkson.edu}
\and
{Carolyn Wendler}}
\date{\today}

\begin{document}
\bibliographystyle{plain}
\maketitle

\begin{abstract}
Classical random walks on well-behaved graphs are rapidly mixing towards the uniform distribution.
Moore and Russell showed that a continuous quantum walk on the hypercube is instantaneously uniform
mixing. We show that the continuous-time quantum walks on other well-behaved graphs do not exhibit
this uniform mixing. We prove that the only graphs amongst balanced complete multipartite graphs 
that have the instantaneous uniform mixing property are 
the complete graphs on two, three and four vertices, and the cycle graph on four vertices.
Our proof exploits the circulant structure of these graphs. Furthermore, we conjecture that most
complete cycles and Cayley graphs lack this mixing property as well.  
\end{abstract}

\section{Introduction}

Two pervasive algorithmic ideas in quantum computation are Quantum Fourier Transform (QFT)
and amplitude amplification (see \cite{cn}).
Most subsequent progress in quantum computing owed much to these two beautiful ideas.
But there are many problems whose characteristics matches neither the QFT nor the amplitude
amplification mold (e.g., the Graph Isomorphism problem). This begs for new additional 
tools to be discovered.

A natural way to discover new quantum algorithmic ideas is to adapt a classical one to the
quantum model. An appealing well-studied classical idea in statistics and computer 
science is the method of random walks \cite{diaconis, motwani}. Recently, the quantum analogue 
of classical random walks has been studied in a flurry of works \cite{fg,cfg,abnvw,aakv,mr,k}. 
The works of Moore and Russell \cite{mr} and Kempe \cite{k} showed faster bounds on instantaneous mixing 
and hitting times for discrete and continuous quantum walks on the hypercube (compared to the classical walk).

The focus of this note is on the {\em continuous}-time quantum walk that was introduced by 
Farhi and Gutmann \cite{fg}. In a subsequent paper, Childs et al. \cite{cfg} gave a simple example where 
the classical and (continuous-time) quantum walks exhibit a different behavior in hitting time 
statistics. 
The goal of this note is to show further qualitative differences (by elementary means) between the two models
in their mixing time behaviors. A main result is that, on the natural class of complete and
balanced complete multipartite graphs, only the complete graphs $K_{n}$ on $n = 2,3,4$ vertices and the 
cycle graph of size four (i.e., $K_{2,2}$) have the instantaneous uniform mixing
property, i.e., there exists a time when the probability distribution function of the quantum walk is
exactly the uniform distribution. This is distinctly different from the classical walk and from the discrete 
quantum walk. The proofs of these exploit heavily the circulant structure of these graphs.

\begin{figure}[h]
\[\begin{xy} /r10mm/:
 ,0		,0*@{*};+/r15mm/*@{*}**@{-}
 ,+/r15mm/      ,{\xypolygon3"B"{~={-30}@{*}}} 
 ,+/r25mm/      ,{\xypolygon4"C"{~={-0}@{*}}} 
 ,+/r25mm/      ,{\xypolygon4"M"{~>{}@{*}}} 
 ,"C1";"C3"**@{-},"C2";"C4"**@{-}
 ,"M1";"M2"**@{-},"M2";"M3"**@{-};"M3";"M4"**@{-},"M4";"M1"**@{-}
 \end{xy}\] 
\caption{The only balanced multipartite graphs with the continuous mixing property.
From left to right: $K_{2}$, $K_{3}=C_{3}$, $K_4$, and $K_{2,2}=C_{4}$.} 
\label{figure:multipartite} 
\end{figure}
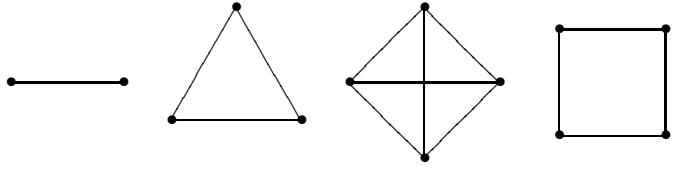

A recent work by Childs et al. \cite{ccdfgs} gave an interesting and powerful algorithmic application of
continuous quantum walks.

\section{Continuous-time quantum walks}

Continuous-time quantum walks was introduced by Farhi and Gutmann \cite{fg} (see also \cite{cfg,mr}). 
Our treatment, though, follow closely the analysis of Moore and Russell \cite{mr} which we review next.
Let $G = (V,E)$ be a simple, undirected, connected, $d$-regular $n$-vertex graph.
Let $A$ be the adjacency matrix of $G$ that is given as
\begin{equation}
	A_{jk} = \left\{\begin{array}{ll}
			1 & \mbox{ if $(j,k) \in E$ } \\
			0 & \mbox{ otherwise }
			\end{array}\right. 
\end{equation}
We define the transition matrix $H = \frac{1}{d}A$ (treated as the Hamiltonian of the quantum system). 
Let the initial amplitude wave function of the particle be $\ket{\psi_0} = \ket{0}$.
The the amplitude wave function at time $t$, is given by Schr\"{o}dinger's equation, as
\begin{equation}
	i\hslash\tderiv{\ket{\psi_{t}}} = H\ket{\psi_{t}}.
\end{equation}
or (assuming from now on that $\hslash = 1$)
\begin{equation}
	\ket{\psi_{t}} = e^{-iHt}\ket{\psi_{0}}.
\end{equation}
This is similar to the model of a continuous-time Markov chain in the classical sense
(see \cite{gs}). It is more natural to deal with the Laplacian of the graph,
which is defined as $L = A - D$, where $D$ is a diagonal matrix with entries 
$D_{jj} = deg(v_{j})$. This is because we can view $L$ as the generator matrix that describes 
an exponential distribution of waiting times at each vertex. 
But on $d$-regular graphs, $D = \frac{1}{d}I$, and since $A$ and $D$ commute, we get
\begin{equation} \label{eqn:phase-factor}
e^{-itL} = e^{-it(A-\frac{1}{d}I)} = e^{-it/d}e^{-itA}
\end{equation}
which introduces an irrelevant phase factor in the wave evolution.

The probability that the particle is at vertex $j$ at time $t$ is given by 
\begin{equation} \label{eqn:collapse}
	P_t(j) = |\braket{j}{\psi_{t}}|^2.
\end{equation}

Since $H$ is Hermitian, the matrix $U_t = e^{-iHt}$ is unitary.
If $(\lambda_j,\ket{z_j})_j$ are the eigenvalue and eigenvector pairs of $H$,
then $(e^{-i\lambda_{j}t},\ket{z_j})_j$ are the eigenvalue and eigenvector pairs of $U_t$.
Because $H$ is symmetric, there is an orthonormal set of eigenvectors, 
say $\{\ket{z_j} : j \in [n]\}$ (i.e., $H$ is unitarily diagonalizable).
So, if $\ket{\psi_0} = \sum_j \alpha_j \ket{z_j}$ then
\begin{equation}
	\ket{\psi_t} = \sum_j \alpha_{j}e^{-i\lambda_{j}t}\ket{z_j}.
\end{equation}
Hence, in order to analyze the behavior of the quantum walk, we follow its wave-like patterns
using the eigenvalues and eigenvectors of the unitary evolution $U_t$. To observe its classical
behavior, we collapse the wave vector into a probability vector using Equation \ref{eqn:collapse}.\\

In the following, we define the notion of (instantaneous) uniform mixing property
of quantum walks (defined earlier in \cite{mr,k}).

\begin{definition}
A graph $G=(V,E)$ has the {\em instantaneous uniform mixing} property if there exists $t \in \Real^{+}$, 
such that $\ket{\psi_{t}}$ of the continuous-time quantum walk on $G$ satisfies 
$P_{t}(j) = 1/|V|$, for all $j \in V$.
\end{definition}

A random walk on $G$ is called {\em simple} if the transition probability matrix is 
$H = \frac{1}{d}A$, where $A$ is the adjacency matrix of the $d$-regular graph $G$. The walk
is called {\em lazy} if, at each step, the walk stays at the current vertex with probability 
$\frac{1}{2}$ and moves according to $H$ with probability $\frac{1}{2}$. 
For continuous-time quantum walks, these two notions are {\em equivalent} modulo a time scaling.
The argument is similar to Equation \ref{eqn:phase-factor} by exploiting the commutativity of
$A$ and $I$ that introduces an irrelevant phase factor in the amplitude expression.

\section{A simple example}

This section describes a single example that illustrates differences between
classical (discrete and continuous) random walk and continuous-time
quantum walk. The example concerns a random walk on $K_{2}$.
Let the Pauli $X$ matrix and the Hadamard matrix $H$ be given as
\begin{equation}
X = \begin{pmatrix} 0 & 1 \\ 1 & 0 \end{pmatrix}, \ \ \ 
H = \frac{1}{\sqrt{2}}\begin{pmatrix} 1 & 1 \\ 1 & -1 \end{pmatrix}.
\end{equation}

The adjacency matrix of $K_{2}$ is given by 
$A = X$. 
Since $K_{2}$ is bipartite, a classical simple random walk starting at the first vertex 
will oscillate between the two states.
So the simple random walk will never reach the uniform distribution.
However, the {\em lazy} random walk, with transition matrix $\frac{1}{2}(I + A)$, will converge
to the uniform distribution after a single step.

Let us consider a continuous-time walk with generator matrix
\begin{equation}
	Q = \begin{pmatrix} -\alpha & \beta \\ \alpha & -\beta \end{pmatrix}.
\end{equation}
This is viewed as a two-state chain
where a transition is made after a waiting time that is exponentially distributed with rates $\alpha$ and
$\beta$ from states $1$ and $2$, respectively (see \cite{gs}, page 156).
Note that $Q$ has eigenvalues $0,-(\alpha + \beta)$ with the following respective eigenvectors 
\begin{equation}
\begin{pmatrix} \beta \\ \alpha \end{pmatrix}, \ \ \ \begin{pmatrix} 1 \\ -1 \end{pmatrix}.
\end{equation}
Let $P(t)$ be the matrix with entries $p_{jk}(t) = \Pr(X(t) = k \ | \ X(0) = j)$ which
describes the evolution of the continuous-time walk after $t$ steps.
It is known that
\begin{equation}
\tderiv{P(t)} = QP(t)
\end{equation}
which implies $P(t) = e^{tQ}$. Note that $Q$ can be diagonalized using the following matrix $B$ 
and its inverse:
\begin{equation}
B = \begin{pmatrix} 1 & \beta \\ -1 & \alpha \end{pmatrix}, \ \ \ \
B^{-1} = \frac{1}{\alpha + \beta} \begin{pmatrix} \alpha & -\beta \\ 1 & 1 \end{pmatrix}.
\end{equation}
Therefore,
\begin{eqnarray*}
P_{t} & = & e^{tQ} = B \begin{pmatrix} e^{-t(\alpha + \beta)} & 0 \\ 0 & 1 \end{pmatrix} B^{-1} \\
	& = & \frac{1}{\alpha + \beta}
		\begin{pmatrix} \alpha e^{-t(\alpha+\beta)}+\beta & \beta(1-e^{-t(\alpha+\beta)}) \\
				\alpha(1-e^{-t(\alpha+\beta)}) & \beta e^{-t(\alpha+\beta)}+\alpha 
		\end{pmatrix}.
\end{eqnarray*}
So, 
the uniform distribution is reached in the limit if and only if
$\alpha = \beta = 1$.

Finally, we consider the continuous-time quantum walk on $K_{2}$. 
The wave amplitude vector at time $t$ is given by
\begin{equation}
\ket{\psi_{t}} 
	= e^{-itA}\ket{0} 
	= \begin{pmatrix} \cos(t) \\ -i\sin(t) \end{pmatrix}.
\end{equation}
\ignore{
\begin{eqnarray*}
\ket{\psi_{t}} 
	& = & e^{itA}\ket{0} = e^{itA}\frac{1}{\sqrt{2}}(H\ket{0} + H\ket{1}) \\
	& = & \frac{1}{2}\begin{pmatrix} 1 + e^{-2it} \\ 1 - e^{-2it} \end{pmatrix} \\
	& = & e^{-it}\begin{pmatrix} \cos(t) \\ i\sin(t) \end{pmatrix}.
\end{eqnarray*}
}
Thus, collapsing this wave vector to its probability form, the expression
\begin{equation}
P_{t} = \begin{pmatrix} \cos^{2}(t) \\ \sin^{2}(t) \end{pmatrix}
\end{equation}
is obtained which reaches the uniform distribution periodically at $t = k\pi + \pi/4$, for $k \in \Int$. 
So, the continuous-time quantum walk on $K_{2}$ has the uniform mixing property. 
We will revisit this analysis under more general conditions in a subsequent section.

\section{Circulant Graphs}

A matrix $A$ is circulant if its $k$-th row is obtained from the
$0$-th row by $k$ consecutive right-rotations. 
A graph $G$ is circulant if its adjacency matrix is a circulant matrix.
Some examples of circulant graphs include complete graphs and full-cycles. 
An important property of circulant matrices is that they are (unitarily) diagonalizable by the
Fourier matrix 
\begin{equation}
	F = \frac{1}{\sqrt{n}}V(\omega),
\end{equation}
where $\omega = e^{2\pi i/n}$ and $V(\omega)$ is the Vandermonde matrix given by
\begin{equation}
	V(\omega) = \left(\begin{array}{lllll}
			1 & 1 & 1 & \ldots & 1 \\
			1 & \omega & \omega^2 & \ldots & \omega^{n-1} \\
			1 & \omega^2 & \omega^4 & \ldots & \omega^{2(n-1)} \\
			\vdots & \vdots & \vdots &  & \vdots \\
			1 & \omega^{n-1} & \omega^{2(n-1)} & \ldots & \omega^{(n-1)^2} 
			\end{array}\right).
\end{equation}
Let the $j$-th column vector of $V(\omega)$ be denoted by $\ket{\omega_j}$, $j=0,\ldots,n-1$.
It is easy to verify that $F$ is unitary, i.e., $F^{-1} = F^{\dagger}$, since the Vandermonde matrix
obeys $V(\omega)^{-1} = V(\omega^{-1})$. 

If $C$ is a circulant matrix whose $0$th column vector is $\ket{f}$ then
\begin{equation} \label{eqn:circ}
	FCF^{\dagger} = diag(V(\omega)\ket{f}).
\end{equation}
This Equation \ref{eqn:circ} shows that the eigenvalues of a circulant matrix can be obtained
by applying the Fourier transform $F$ to the first column vector of $C$.

\section{Classical versus quantum walks} \label{section:complete}

\subsection{The complete graph $K_n$}

The adjacency matrix of $K_n$ is $A = J_n - I_n$, where $J_n$ is the all-one $n \times n$ matrix
and $I_n$ is the $n \times n$ identity matrix.
The eigenvalues of $\frac{1}{n-1}A$ are $1$ (once) and $-\frac{1}{n-1}$ ($n-1$ times).
Using the orthonormal eigenvectors $\ket{F_j} \dotdef \frac{1}{\sqrt{n}}\ket{\omega_j}$ 
(the columns of the Fourier matrix $F$), 
the initial amplitude vector is $\ket{\psi_0} = \frac{1}{n}\sum_{j} \ket{\omega_j}$.
We get
\begin{equation}
\braket{j}{\psi_t} = 
	\left\{
	\begin{array}{ll}
	-\frac{2}{n}e^{\frac{-it(n-2)}{2(n-1)}}sin\left(\frac{tn}{2(n-1)}\right) 
		& \mbox{ if $j \neq 0$ } \\
	\frac{1}{n}(e^{-it} + (n-1)e^{it/(n-1)})
		& \mbox{ if $j = 0$ }
	\end{array}\right.
\end{equation}
Thus
\begin{equation}
	P_t(j) = 
	\left\{
	\begin{array}{ll}
	\frac{4}{n^2}sin^2\left(\frac{tn}{2(n-1)}\right).
		& \mbox{ if $j \neq 0$ } \\
	1 - \frac{4(n-1)}{n^2}sin^2\left(\frac{tn}{2(n-1)}\right)
		& \mbox{ if $j = 0$ }
	\end{array}\right.
\end{equation}
Notice that to achieve uniformity, for all $j \neq 0$, $P_t(0) = P_t(j)$ or
\begin{equation}
	\frac{4}{n}sin^2\left(\frac{tn}{2(n-1)}\right) = 1.
\end{equation}
Uniformity is possible only if $n = 2,3,4$.
In contrast, a classical walk never achieves uniform on $K_2$ but converges to uniform 
for all $K_n$, $n > 2$. This is summarized in the following theorem.

\vspace*{10pt}
\begin{theorem} \label{thm:complete}
No complete graph, except for $K_2$, $K_3$, and $K_4$, has
the uniform mixing property under the continuous-time quantum model.
\end{theorem}

\subsection{The balanced complete multipartite graphs}

Let $G$ be a complete $a$-partite graph where each partition has $b > 1$ vertices 
(the case $b = 1$ is the complete graph case).
Let $A$ be the normalized adjacency matrix of $G$. 
Note that $A$ is given by
\begin{equation}
A = \frac{1}{a-1}K_{a} \otimes \frac{1}{b}J_{b}.
\end{equation}
Using the circulant structure of both matrices, it is clear that
the normalized eigenvalues of $K_{a}$ are $1$ (once) and $-\frac{1}{a-1}$ (with multiplicity $a-1$)
and that the normalized eigenvalues of $J_{b}$ are $1$ (once) and $0$ (with multiplicity $b-1$).
Let $\alpha = e^{\frac{2\pi i}{a}}$ be the principal $a$-th root of unity and let
$\beta = e^{\frac{2\pi i}{b}}$ be the principal $b$-th root of unity. 
Both matrices have the columns of the Fourier matrix of dimensions $a$ and $b$, respectively, 
as their orthonormal set of eigenvectors. 
So let $\{\ket{\alpha_j}\}_{j=0}^{a-1}$ and $\{\ket{\beta_k}\}_{k=0}^{b-1}$ 
be the orthogonal sets of eigenvectors of $K_a$ and $J_b$, respectively.
Note that $\ket{\alpha_0} = \ket{1_a}$ and $\ket{\beta_0} = \ket{1_b}$.

The eigenvalues of $A$ is $\lambda_0 = 1$ (once) with the all-one eigenvector 
\begin{equation}
	\ket{1_a} \otimes \ket{1_b},
\end{equation}
$\lambda_1 = -\frac{1}{a-1}$ (with multiplicity $a-1$) with the eigenvectors (with lengths $ab$)
\begin{equation}
	\ket{\alpha_j} \otimes \ket{1_b}, \ \ \ j = 1,2,\ldots,a-1.
\end{equation}
and $\lambda_2 = 0$ (with multiplicity $a(b-1)$) with eigenvectors (with lengths $ab$)
\begin{equation}
	\ket{\alpha_j} \otimes \ket{\beta_k}, \ \ \ 1 \le j < a, \ 1 \le k < b.
\end{equation}
Thus, the wave amplitude function $\ket{\psi_{t}}$ is given by
\begin{equation}
\ket{\psi_t} = 
	\frac{1}{ab}\left[
	e^{-it} \ket{1_a} \otimes \ket{1_b} 
	+
	e^{\frac{it}{a-1}}
	\sum_{j=1}^{a-1}
	\ket{\alpha_j} \otimes \ket{1_b} 
	+ 
	\sum_{k=1}^{b-1}
	\sum_{j=1}^{a-1} \ket{\alpha_j} \otimes \ket{\beta_k} 
	\right]
\end{equation}
which equals to
\begin{equation}
\ket{\psi_t} = 
	\frac{1}{ab}\left[
	e^{-it}\left(\begin{array}{c} \ket{1_b} \\ \ket{1_b} \\ \vdots \\ \ket{1_b} \end{array}\right) 
	+
	e^{\frac{it}{a-1}}
	\left(\begin{array}{c} \ket{a}-\ket{1_b} \\ -\ket{1_b} \\ \vdots \\ -\ket{1_b} \end{array}\right) 
	+
	\left(\begin{array}{c} a\sum_{k=1}^{b-1} \ket{\beta_k} \\ \ket{0_b} \\ \vdots \\ \ket{0_b} \end{array}\right)
	\right].
\end{equation}
The wave amplitude expressions are
\begin{equation}
\braket{j}{\psi_t} = 
	\left\{
	\begin{array}{ll}
	\frac{1}{ab}(e^{-it} + e^{\frac{it}{a-1}}(a-1) + a(b-1))
		& \mbox{ if $j = 0$ } \\
	\frac{1}{ab}(e^{-it} + e^{\frac{it}{a-1}}(a-1) - a)
		& \mbox{ if $1 \le j < b$ } \\
	\frac{1}{ab}(e^{-it} - e^{\frac{it}{a-1}}) 
		& \mbox{ otherwise }
	\end{array}\right.
\end{equation}
Forcing the uniformity on the last type, for $j \ge b$, one obtains
\begin{equation}
P_t(j) = \frac{4}{(ab)^2}\sin^{2}\left[\frac{ta}{2(a-1)}\right] = \frac{1}{ab},
\end{equation}
which yields the condition
$\sin^{2}(ta/[2(a-1)]) = \frac{ab}{4}$
or
$1 \le ab \le 4$.
There are only three {\em legal} (integral) cases, namely,
$a=1,b=4$ (or $\overline{K_{4}}$), $a=b=2$ (or $K_{2,2}$), and $a=4,b=1$ (or $K_{4}$).
The case for the empty graph $\overline{K_{4}}$ is obvious and we know from Theorem
\ref{thm:complete} that $K_{4}$ has uniform mixing. For the case of $K_{2,2}$,
the three amplitude expressions are
$\frac{1}{4}(e^{it}+e^{-it}+2) = \frac{1}{2}(\cos(t)+1)$,
$\frac{1}{4}(e^{it}+e^{-it}-2) = \frac{1}{2}(\cos(t)-1)$, 
and
$\frac{i}{2}\sin(t)$. For $t$ being odd multiples of $\pi/2$, their probability forms
achieve uniformity. So $K_{2,2}$ has uniform mixing.

\begin{theorem} \label{thm:multipartite}
No balanced complete multipartite graph, except for $K_{2,2}$, has
the uniform mixing property under the continuous-time quantum walk model.
\end{theorem}

\section{Other graphs}

\subsection{The complete cycle $C_n$}

\par\noindent
The adjacency matrix $A$ of $C_n$ is given by
\begin{equation}
A = \left(\begin{array}{cccccc} 
	  0 & 1 & 0 & \ldots & 0 & 1 \\ 
	  1 & 0 & 1 & \ldots & 0 & 0 \\ 
	  0 & 1 & 0 & \ldots & 0 & 0 \\ 
	  \vdots & \vdots & \vdots & \vdots & & \vdots \\ 
	  1 & 0 & 0 & \ldots & 1 & 0 
	  \end{array}\right).
\end{equation}
Let $H = \frac{1}{2}A$ be the normalized adjacency matrix of the full-cycle $C_n$ on $n$ vertices.
Let $\omega_{j} = \omega^{j} = e^{2\pi ij/n}$, for $j = 0,1,\ldots,n-1$.
Using the properties of circulants, the eigenvalues of $H$ is given by
\begin{equation}
	\lambda_{j} = \frac{1}{2}(\omega_{j} + \omega_{j}^{n-1}) = \cos(2\pi j/n).
\end{equation}
The wave equation gives us
\begin{equation}
\ket{\psi_t} = U_t \ket{\psi_0} = \frac{U_t}{n}\sum_{j}\ket{\omega_j} 
	= \frac{1}{n}\sum_{j=0}^{n-1} e^{-i\lambda_{j}t}\ket{\omega_j} 
	= \frac{1}{n}\sum_{j=0}^{n-1}e^{-it\cos(2\pi j/n)} \ket{\omega_j}
\end{equation} 
Then 
\begin{equation} 
	\braket{k}{\psi_t} =
	\frac{1}{n}\sum_{j=0}^{n-1} e^{-it\cos(2\pi j/n)}\omega_{j}^{k}. 
\end{equation} 
This is a complicated nested exponential sum\footnote{An alternative expression 
is $\braket{k}{\psi_{t}} = \sum_{\nu \equiv \pm k\pmod{N}} (-i)^{\nu} J_{\nu}(t)$,
where $J_{\nu}(t)$ is the Bessel function \cite{bbt}.}.
Despite the fact that $C_{3}$ (which is equivalent to $K_{3}$)
and $C_{4}$ (which is equivalent to $K_{2,2}$) both have the uniform mixing property,
the general case of $C_{n}$, $n > 4$, remains intractable. 

\begin{conjecture}
No complete cycle $C_{n}$, except for $C_{3}, C_{4}$, has the instantaneous uniform mixing property
under the continuous-time quantum walk model.
\end{conjecture}

Note that $C_{n}$ has the uniform mixing property in the discrete quantum walk model \cite{aakv}.

\subsection{The Cayley graph $S_n$}

\par\noindent
Let $S_n$ be the symmetric group on $n$ elements (the group of all permutations on $[n]$)
and let $T_n$ be the set of transpositions on $[n]$. 
The Cayley graph $X_{n}(S_n,T_n)$ is defined on the vertex set $S_n$
and $(\pi,\tau \pi)$ is an edge, for all $\pi \in S_n$ and $\tau \in T_n$;
it is a bipartite, connected $\binom{n}{2}$-regular graph on $n!$ vertices.
Consider the simplest case of $S_3$ with elements
$\{e,(1 \ 2),(1 \ 3),(2 \ 3),(1 \ 2 \ 3),(1 \ 3 \ 2)\}$.
This Cayley graph on $S_3$ is isomorphic to $K_{3,3}$.
By Theorem \ref{thm:multipartite}, the continuous-time quantum walk on it is
not uniform mixing. We have verified that the same is true of $S_{4}$. 

\begin{claim}
The Cayley graph of $S_4$ is not instantaneously uniform mixing.
\end{claim}

We conjecture that this phenomenon holds for all $n > 4$. 

\begin{conjecture}
No Cayley graph $X_{n}$, except for $X_{3}$, has the instantaneous uniform mixing property
under the continuous-time quantum walk model.
\end{conjecture}

It is likely that tools from group representations and characters might
settle this question. This is because the characters of $S_n$ yield information about the 
eigenvalues of the Cayley graph. Given this, one could build the set of orthonormal eigenvectors. 
Unfortunately, there is no known explicit formula that gives all the characters of $S_n$; 
only estimates are known for the general case.

\section*{Acknowledgements}

\noindent
This work is supported in part by the National Science Foundation under grants DMR-0121146 and
DMS-0097113. We thank Pedro Tejada for discussions.

\end{document}